\documentclass[preprint,12pt]{elsarticle}

\usepackage{amsmath,amssymb,bm,color,multirow,booktabs}
\usepackage{caption}
\usepackage{subcaption}
\usepackage{hyperref}
\usepackage{wrapfig}

\newcommand{\Matrix}[1]{\bm{#1}}  % for matrices
\newcommand{\refFig}[1]{Figure~\ref{#1}}           % figure
\newcommand{\refSec}[1]{Section~\ref{#1}}        % section
\newcommand{\Revised}[1]{\textcolor{black}{#1}} % Used to highlight the changes made to the original manuscript

\usepackage{eso-pic} % for arXiv note
\usepackage{xcolor}

\journal{}

\makeatletter
\def\ps@pprintTitle{%
  \let\@oddhead\@empty
  \let\@evenhead\@empty
  \def\@oddfoot{\reset@font\hfil\thepage\hfil}
  \let\@evenfoot\@oddfoot
}
\makeatother
\begin{document}

% ----------arXiv-------------------------
\AddToShipoutPicture*{%
  \AtPageUpperLeft{%
    \setlength\unitlength{1cm}%
    \put(0,-1.4){\begin{minipage}[c]{\paperwidth}
    \footnotesize\centering\textcolor{black!50}{%
    This is a preprint of an article published in \emph{Mechatronics}\\
    The final authenticated version is available online at:} \ \textcolor{blue!60}{\url{doi.org/10.1016/j.mechatronics.2023.103114}}\\
    \footnotesize\centering\textcolor{black!50}{%
    \copyright 2023. Licensed under the Creative Commons CC-BY-NC-ND 4.0 license}%
    \end{minipage}}%
  }
}
% ---------------------------------------

\begin{frontmatter}



\title{Active disturbance rejection control for unmanned tracked vehicles in leader-follower scenarios: discrete-time implementation and field test validation}


\author[inst1]{Salem-Bilal Amokrane}
\author[inst1]{Mohammed Zouaoui Laidouni}
\author[inst1]{Touati Adli}

\affiliation[inst1]{organization={Military Academy, University of Defence},%Department and Organization
            %addressline={}, 
            city={Belgrade},
            %postcode={}, 
            %state={},
            country={Serbia}}

\author[inst2]{Rafal Madonski\corref{cor1}}
\ead{rmadonski@polsl.pl}
\author[inst1]{Momir Stanković}

\affiliation[inst2]{organization={Faculty of Automatic Control, Electronics and Computer Science,\\ Silesian University of Technology},%Department and Organization
            %addressline={ul. Alademicka 16}, 
            city={Gliwice},
            %postcode={44-100}, 
            %state={},
            country={Poland}}
            
\cortext[cor1]{Corresponding author}

\begin{abstract}
This paper presents a systematic design of an active disturbance rejection control (ADRC) system for unmanned tracked vehicles (UTVs) in leader-follow formation. Two ADRC controllers are designed for the lateral and the longitudinal channels of the UTV based on control errors in the cross-track and the along-track directions. Through simulations, the proposed ADRC approach is first shown to outperform the conventional PI/PID controllers in scenarios involving sudden changes in the leader motion dynamics, slippage disturbances, and measurement noise. Then, a comprehensive experimental validation of the proposed leader-follower control is performed using a laboratory UTV equipped with a camera and laser sensors (to enable the calculation of error signals). In order to provide more effective interaction between the human (leader) and the UTV (follower) during the leader-follower task, a camera-based subsystem for human pose recognition is developed and deployed. Finally, the experimental results obtained outdoors demonstrate that the proposed ADRC-based leader-follower UTV control system achieves high tracking capabilities, robustness against slippage disturbances, and adaptability to changing environmental conditions.\\
\vspace{0.01em}\\
\textit{\Revised{Supplementary  material}}: \url{https://youtu.be/rQmnVLZQc8o}
\end{abstract}

\begin{keyword}
unmanned tracked vehicle (UTV) \sep active disturbance rejection control (ADRC) \sep leader-follower control \sep PI/PID
\end{keyword}

\end{frontmatter}


\section{Introduction}
\label{sec:Section1}

%%%%%%%%%%%%%%%%%%%%%%%%%%%%%%%%%%%%%%
%%%%%%%%%%%%%%%%%%%%%%%%%%%%%%%%%%%%%%
%%%%%%%%%%%%%%%%%%%%%%%%%%%%%%%%%%%%%%

In recent years, the concept of leader-follower control in unmanned vehicles (UVs) has gained significant attention in various applications, including surveillance, logistics, precise agriculture, and search and rescue missions \cite{eskandarian2019research}. By incorporating advanced sensing technologies based on optical cameras and/or laser-based sensors, coupled with robust control algorithms, UV can accurately perceive and autonomously track a designated leader, typically a human operator or a leading vehicle, while maintaining a safe distance and matching its movements 
\cite{verma2018vehicle}, \cite{kobilarov2006people}. The camera-based leader-follower solutions extract valuable visual information to detect and track the leader, using object recognition algorithms \cite{ibrahim2020vision} \cite{quadrafollowMe,monocameraFollowMe} or specific camera systems \cite{kang2019adaptive}, enabling precise leader-vehicle interaction and tracking performances. On the other hand, utilizing laser sensors \cite{lidar_follow}, \cite{kawarazaki2015development}, in the leader-follower control, allows the ability to operate in various lighting conditions, robustness against environmental factors such as shadows or changes in ambient light, and the capability to accurately perceive both stationary and moving leader.

However, regardless of the chosen sensor technology for leader detection and tracking error estimation, the instantaneous changes in the leader motion direction and the relatively short distances between the leader and the UV require high dynamics in the UV  motion. Additionally, considering the structural limitations of UVs, the influence of the vehicle's unmodeled dynamics, sensor measurement noise, and slippage disturbances, it becomes evident that from a control design perspective, the leader-follower task is a highly challenging problem. Addressing these demanding requirements necessitates the development of an advanced control technique that can handle rapid changes in motion, adapt to varying environmental conditions, and robustly respond to drive wheel/track slippage. Actually, the designed control system should provide a fast response of the vehicle with a minimal turning radius, which, in turn, can cause wheel or track slippage disturbance \cite{janarthanan2012longitudinal}. It introduces uncertainties and affects the vehicle's ability to accurately track the leader and compromises the vehicle stability performance. 
 
In the current literature, various control strategies have been analyzed to ensure the appropriate execution of the desired UV control tasks, but generally, all approaches can be classified into four main groups: classical proportional integral derivative (PID) control, intelligent control, modern control, and observer-based control. To this point, in \cite{cao2022design}, two PID controllers were utilized in order to control the steering wheel and vehicle speed, while \cite{zou2018dynamic} proposes a back-stepping UV wheel torque control methodology based on a modified PID algorithm. Intelligent control methods, such as fuzzy logic and neural networks, are often combined with classical controllers to achieve high tracking performance in challenging terrains \cite{li2020path,dai2018trajectory}. On the other hand, modern solutions rely on techniques like model predictive control \cite{mitsuhashi2019autonomous}, adaptive control \cite{Cui2014,gonzalez2010adaptive}, passivity-based control \cite{Li2022Tele,LiTAC}, sliding mode control \cite{sabiha2022ros}, or observer-based control \cite{burke2012path,hiramatsu2019path} which involves estimating various parameters and disturbances to improve control performance without significantly increasing the complexity of the model.

Overall, these researches showcase different approaches to control UVs, highlighting the trade-offs between performance, robustness, computational requirements, and complexity. While classical PID controllers are commonly employed and possess a solid theoretical foundation, they exhibit inherent limitations that can affect their overall performance. One such limitation is their reliance on fixed tuning parameters, which are usually determined based on a particular operating point or assumed system dynamics. As a result, PID controllers may struggle to adapt to varying UV operating conditions or changes in system dynamics, leading to degradation of the control performances. The intelligent control methods typically have a limited hardware implementation due to algorithms' complexity and high computational requirements, while the modern control methodology requires precise modeling of UV motion and slippage disturbance. Finally, observer-based techniques contribute to improved system performance, robustness to slippage disturbances and measurement noise, and adaptability to changing operating conditions. However, sensitivity to model uncertainties and unmodelled dynamics, as well as computational complexity, especially for UV systems with high-dimensional state spaces, can limit their implementation. 

In order to facilitate the design of UV leader-follower control with minimal requirements for knowledge of the UV model and slippage dynamics, this paper proposes the utilization of active disturbance rejection control (ADRC) methodology, which can be considered as a combination of the classical and observer-based control techniques \cite{gao2006active,zhang2021overview}. In the context of UV control, the advantages of the ADRC concept were recently analyzed in UV path following \cite{sen2019active} and trajectory tracking \cite{wang2021composite,wang2023switching,neria2023} tasks, where it was shown that ADRC allows effective control in dynamic and uncertain environments, where disturbances such as slippage dynamics, terrain variations, or sensor noise affect the vehicle's behavior. The ADRC's ability to estimate system disturbances by a dedicated extended state observer (ESO) and then compensate for these disturbances, enables precise and responsive control actions, enhancing the overall performance and stability of the UV in different working conditions.

This research extends the application of the ADRC methodology to a specific and challenging task of the leader-follower control of UVs, which involves dealing with the rapidly changing dynamics of the leader's movement and potential measurement noise in the signals that facilitate the interaction between the vehicle and leader. The contributions of this paper can be categorized into academic and practical engineering aspects. From an academic perspective, this work presents a definition of the leader-follower control problem for specific unmanned tracked vehicles (UTV) and the systematic design of ADRC-based controllers for the lateral and longitudinal control channels. On the practical front, the research proposes a comprehensive experimental leader-follower control setup using laboratory UTV with a camera and laser sensors-based algorithms for track error estimation and a human leader pose recognition algorithm, which enables more effective human-vehicle interaction. The achieved simulation results have demonstrated that the proposed solution provides significantly better performance compared to the industrial PI/PID controller, while the experimental verification has confirmed its validity and effectiveness in real leader-follower scenarios. It has been observed that the UV exhibits enhanced tracking capabilities, robustness to slippage disturbances, and adaptability to changing environmental conditions. 

The rest of the paper is organized as follows. Section~\ref{sec:Section2} defines the UTV leader-follower control problem. Then, in Section~\ref{sec:Section3}, the proposed ADRC design for lateral and longitudinal control channels is outlined. Next, Section~\ref{sec:Section4} shows the simulation results. The experimental validation of the proposed ADRC-based UTV leader-follower control task is discussed in Section~\ref{sec:Section5}. Finally, Section~\ref{sec:Section6} concludes the paper.

%%%%%%%%%%%%%%%%%%%%%%%%%%%%%%%%%%%%%%
%%%%%%%%%%%%%%%%%%%%%%%%%%%%%%%%%%%%%%
%%%%%%%%%%%%%%%%%%%%%%%%%%%%%%%%%%%%%%

\section{Definition of UTV leader-follower control problem}
\label{sec:Section2}

%%%%%%%%%%%%%%%%%%%%%%%%%%%%%%%%%%%%%%
%%%%%%%%%%%%%%%%%%%%%%%%%%%%%%%%%%%%%%
%%%%%%%%%%%%%%%%%%%%%%%%%%%%%%%%%%%%%%

Let us consider a disturbance-free dynamic model of UTV:
\begin{equation}
    \begin{aligned}
        v(t) &= \frac{r}{2}\cdot\left[\Omega_\mathrm{R}(t)+\Omega_\mathrm{L}(t)\right], \\
        \dot{\theta}(t) &= \frac{r}{B}\cdot\left[\Omega_\mathrm{R}(t) - \Omega_\mathrm{L}(t)\right],
        \label{eqn:LongitudinalAngular_vehicle_velocity_ideal}
    \end{aligned}
\end{equation}
where $v(t)$ is the vehicle longitudinal velocity, $\theta(t)$ is the vehicle angular course in the inertial coordinate system, $B$ represents the normal distance between the right and left track, $r$ is the radius of the track drive wheel, $\Omega_\mathrm{R}(t)$ and $\Omega_\mathrm{L}(t)$ are the angular velocities of the right and left track drive wheel, respectively. 

In the presence of track slippage disturbances, \eqref{eqn:LongitudinalAngular_vehicle_velocity_ideal} can be modified to more accurately represent the UTV dynamics:
\begin{equation}
    \begin{aligned}
        v(t)+v_\mathrm{d}(t) &= \frac{r}{2}\cdot\left[a_\mathrm{R}(t)\cdot\Omega_\mathrm{R}(t) + a_\mathrm{L}(t)\cdot\Omega_\mathrm{L}(t)\right], \\
        \dot{\theta}(t)+\dot{\theta}_\mathrm{d}(t) &= \frac{r}{B}\cdot\left[a_\mathrm{R}(t)\cdot\Omega_\mathrm{R}(t) - a_\mathrm{L}(t)\cdot\Omega_\mathrm{L}(t)\right], \label{LongitudinalAngular_vehicle_velocity_real}
    \end{aligned}
\end{equation}
where $v_\mathrm{d}(t)$ and $\theta_\mathrm{d}(t)$ are unknown slippage disturbances in the longitudinal velocity and the angular course dynamics, respectively, caused by unknown friction coefficients of the right and left tracks $a_\mathrm{R}(t)$ and $a_\mathrm{L}(t)$, respectively, which are in range $[0,1]$, where the value 1 corresponds to a case without slippage disturbance, and value 0 denotes a case with the complete slippage of the track.

\begin{figure}[htb!]
    \centering
    \includegraphics[width=0.75\columnwidth]{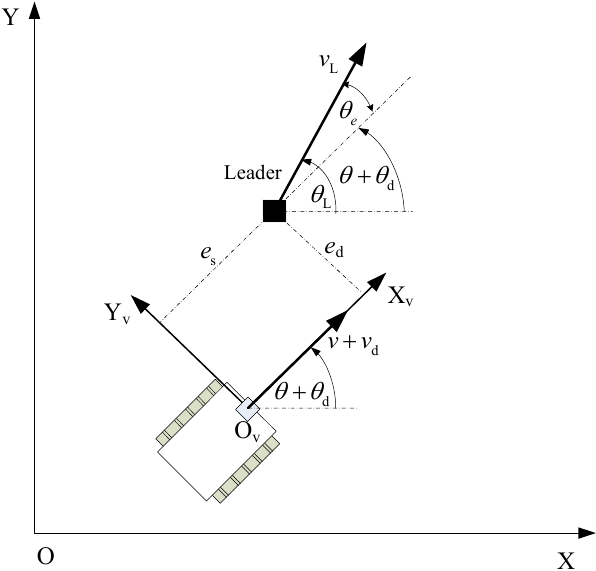} 
     \caption{The considered leader-follower UTV control problem (with assumed notation).}     
     \label{fig:Follow me problem}
\end{figure}

To analyze the UTV leader-follower control problem, let us consider its graphical representation shown in Figure~\ref{fig:Follow me problem}, where the UTV is tasked to follow a leader, which is denoted with its angular orientation $\theta_\mathrm{L}(t)$ in the inertial coordinate system ($\mathrm{X}$, $\mathrm{Y}$) and longitudinal velocity $v_\mathrm{L}(t)$.  

By defining the errors between the leader and the follower in the vehicle coordinate system (X$_\mathrm{v}$, Y$_\mathrm{v}$), the leader-follower UTV control task can be defined as two simultaneous problems: regulation of the cross-track error $e_\mathrm{d}(t)$ (i.e. lateral control channel) and tracking the desired value of the along-track error $e_\mathrm{s}(t)$ (i.e. longitudinal control channel). The dynamics of both control channels are derived next.  

%%%%%%%%%%%%%%%%%%%%%%%%%%%%%%%%%%%%%%
\subsection{Lateral control channel}
\label{lateral_channel}
%%%%%%%%%%%%%%%%%%%%%%%%%%%%%%%%%%%%%%

From Figure~\ref{fig:Follow me problem}, showing the considered leader-follower control problem, the dynamics of the cross-track error yields:
\begin{equation}
    \dot{e}_\mathrm{d}(t) = v_\mathrm{L}(t) \cdot \sin{\theta_\mathrm{e}(t)}, \label{eqn:cross-track_error}
\end{equation}
where $\theta_\mathrm{e}(t)$ is the course angle error between leader and UTV:
 \begin{equation}
 \theta_\mathrm{e}(t)=\theta_\mathrm{L}(t)-\left[\theta(t)+\theta_\mathrm{d}(t)\right].
 \label{eqn:course_angle_error}
 \end{equation}
By substituting \eqref{eqn:course_angle_error} in \eqref{eqn:cross-track_error}, the lateral control channel dynamic can be obtained as:
\begin{align}
    \Ddot{e}_\mathrm{d}(t) &= -v_\mathrm{L}(t) \cdot \cos{\theta_\mathrm{e}(t)} \cdot \dot{\theta}(t) + \dot{v}_\mathrm{L}(t) \cdot \sin\theta_\mathrm{e}(t) \notag \\
    &+ v_\mathrm{L}(t) \cdot \cos{\theta_\mathrm{e}(t)} \cdot \left[\dot{\theta}_\mathrm{L}(t)-\dot{\theta}_\mathrm{d}(t)\right], 
    \label{eqn:lateral_channel}
\end{align}
where regulation of $e_\mathrm{d}(t)$ should be achieved by the appropriate control input $\dot{\theta}(t)$,in the presence of disturbances caused by the leader course angle dynamic $\dot{\theta}_\mathrm{L}(t)$ and slippage dynamic in lateral control channel $\dot{\theta}_\mathrm{d}(t)$. 

%%%%%%%%%%%%%%%%%%%%%%%%%%%%%%%%%%%%%%
\subsection{Longitudinal control channel}
%%%%%%%%%%%%%%%%%%%%%%%%%%%%%%%%%%%%%%

By analyzing Figure~\ref{fig:Follow me problem}, the dynamics of the along-track error can be written as:
\begin{equation}
    \dot{e}_\mathrm{s}(t) = v_{\mathrm{L}}(t) \cdot \cos{\theta_\mathrm{e}(t)} - \left[v(t)+v_\mathrm{d}(t)\right].
    \label{eqn:Along-track_error}
\end{equation}
A longitudinal channel control task can be defined as a tracking problem, where the controlled output ${e}_\mathrm{s}(t)$ should track the desired reference value ${e}_\mathrm{s(ref)}(t)$ with the use of control input $v(t)$, even in the presence of variable leader longitudinal velocity $v_{\mathrm{L}}(t)$ and longitudinal slippage disturbance $v_{\mathrm{d}}(t)$.

    %%%%%%%%%%%%%%%%%%%%%%%%%%%%%%%%%%%%%%
%%%%%%%%%%%%%%%%%%%%%%%%%%%%%%%%%%%%%%
%%%%%%%%%%%%%%%%%%%%%%%%%%%%%%%%%%%%%%

\section{Main result: ADRC-based leader-follower control design}
\label{sec:Section3}

%%%%%%%%%%%%%%%%%%%%%%%%%%%%%%%%%%%%%%
%%%%%%%%%%%%%%%%%%%%%%%%%%%%%%%%%%%%%%
%%%%%%%%%%%%%%%%%%%%%%%%%%%%%%%%%%%%%%

%%%%%%%%%%%%%%%%%%%%%%%%%%%%%%%%%%%%%%
\subsection{Lateral ADRC controller design}
\label{Section:Lateral ADRC design}
%%%%%%%%%%%%%%%%%%%%%%%%%%%%%%%%%%%%%%

Let us represent the lateral control error dynamic \eqref{eqn:lateral_channel} in a more compact form as:  
\begin{equation}
    \ddot{e}_\mathrm{d}(t) = b(t)  \cdot  \dot{\theta}(t)+ d_\mathrm{l}(t),   \label{eqn:lateral_control_system0}
\end{equation}
where $b(t) = -{v}_\mathrm{L}(t) \cdot \cos(\theta_\mathrm{e}(t))$ is the system gain and:
\begin{equation}
    d_\mathrm{l}(t) = \dot{v}_\mathrm{L}(t) \cdot \sin\theta_\mathrm{e}(t) + v_\mathrm{L}(t) \cdot \cos{\theta_\mathrm{e}(t)} \cdot \left[\dot{\theta}_\mathrm{L}(t)-\dot{\theta}_\mathrm{d}(t)\right],
\end{equation}
is the external disturbance in the channel. By applying the ADRC framework, the \eqref{eqn:lateral_control_system0} can rewritten as:
\begin{equation}
    \Ddot{e}_\mathrm{d}(t)=b_0\cdot\dot{\theta}(t)+ f_\mathrm{l}(t),   \label{eqn:lateral_control_system}
\end{equation}
where $b_0=-\bar{v}_\mathrm{L}$ is the assumed (expected) value of the leader longitudinal velocity, and:
\begin{equation}
    f_\mathrm{l}(t)=d_\mathrm{l}(t)+\left[\bar{v}_\mathrm{L}-b(t)\right]\cdot\dot{\theta}(t),
\end{equation}
denotes the total (lumped) disturbance in the lateral channel.

In that manner, \eqref{eqn:lateral_control_system} can be represented in the extended state-space form as:
\begin{align}
    \begin{bmatrix} \dot{e}_\mathrm{d}(t)\\ \ddot{e}_\mathrm{d}(t)\\ \dot{f}_\mathrm{l}(t)\end{bmatrix}&=
    \underbrace{\begin{bmatrix} 0 & 1 & 0\\ 0 & 0 & 1\\ 0 & 0 & 0\end{bmatrix}}_{\Matrix{A}_\mathrm{l}}\cdot\begin{bmatrix} e_\mathrm{d}(t)\\ \dot{e}_\mathrm{d}(t)\\ f_\mathrm{l}(t)\end{bmatrix}+\underbrace{\begin{bmatrix} 0\\b_0\\0\end{bmatrix}}_{\Matrix{B}_\mathrm{l}}\cdot\dot{\theta}(t)+\begin{bmatrix} 0\\0\\ 1\end{bmatrix}\cdot\dot{f}_1(t).
    \label{eqn:extended_state_space_model_lateral}
\end{align}

The states of system \eqref{eqn:extended_state_space_model_lateral} can be estimated based on measured system output $e_\mathrm{d}(t)$ using an extended state observer (ESO):
\begin{align}
    \begin{bmatrix} \hat{\dot{e}}_\mathrm{d}(t)\\ \hat{\ddot{e}}_\mathrm{d}(t)\\ \hat{\dot{f}}_1(t)\end{bmatrix}=\Matrix{A}_\mathrm{l} \cdot \begin{bmatrix} \hat{e}_\mathrm{d}(t)\\ \hat{\dot{e}}_\mathrm{d}(t)\\ \hat{f}_1(t)\end{bmatrix}+\Matrix{B}_\mathrm{l}\cdot\dot\theta(t)+\begin{bmatrix} l_\mathrm{1l}\\l_\mathrm{2l}\\ l_\mathrm{3l}\end{bmatrix}\cdot\left[e_\mathrm{d}(t)-\hat{e}_\mathrm{d}(t)\right],
    \label{eqn:Lateral_ESO}
\end{align}
where $l_\mathrm{1l}$, $l_\mathrm{2l}$, and $l_\mathrm{3l}$ are observer gains.
Based on the estimated system states, the control law for lateral channel is defined as:
\begin{equation}
   \dot\theta(t)=\frac{1}{b_0} \cdot \left[-k_\mathrm{1l}\cdot \hat{e}_\mathrm{d}(t)-k_\mathrm{2l}\cdot \hat{\dot{e}}_\mathrm{d}(t)-\hat{f}_1(t)\right], 
   \label{eqn:Lateral_ADRC_control_law}
\end{equation}
where $k_\mathrm{1l}$ and $k_\mathrm{2l}$ are controller tuning parameters.

It should be noted that by assuming $f_\mathrm{l}(t)\approx\hat{f}_\mathrm{l}(t)$,  $\dot{e}_\mathrm{d}(t)\approx\hat{\dot{e}}_\mathrm{d}(t)$, $e_\mathrm{d}(t)\approx\hat{e}_\mathrm{d}(t)$ and substituting \eqref{eqn:Lateral_ADRC_control_law} in \eqref{eqn:lateral_control_system}, the lateral control channel dynamics is reduced to:
\begin{align}
\ddot{e}_\mathrm{d}(t)+k_\mathrm{2l}\cdot \dot{e}_\mathrm{d}(t)+k_\mathrm{1l}\cdot e_\mathrm{d}(t)\approx0,
\label{eqn:Lateral_chanell_approximation}
\end{align}
which can be tuned by the appropriate selection of  parameters $k_\mathrm{1l}$ and  $k_\mathrm{2l}$.

In this research, the controller parameters are tuned according to the \emph{bandwidth parameterization} method \cite{1242516} by placing closed-loop controller poles at common location $\lambda = -\omega_\mathrm{CL_l}$:
\begin{align}
    \left( \lambda + \omega_\mathrm{CL_l} \right)^2
    \stackrel{!}{=} \lambda^{2}  + k_\mathrm{2l}\cdot  \lambda + k_\mathrm{1l}, 
    \label{tuning_lateral_controller_K}
\end{align}
where $\omega_\mathrm{CL_l}$ is the desired closed-loop lateral control system bandwidth. Similarly, the ESO \eqref{eqn:Lateral_ESO} gains are tuned by placing its poles at common location  $\lambda = -\omega_\mathrm{ESO_l}$, i.e.:
\begin{align}
    \left( \lambda + \omega_\mathrm{ESO_l} \right)^3
    \stackrel{!}{=} \lambda^{3}  + l_\mathrm{1l}\cdot  \lambda^{2} + l_\mathrm{2l}\cdot  \lambda+l_\mathrm{3l} , 
    \label{tuning_lateral_controller_lambda}
\end{align}
where $\omega_\mathrm{ESO_l}$ is the desired ESO bandwidth. 

%%%%%%%%%%%%%%%%%%%%%%%%%%%%%%%%%%%%%%
\subsection{Longitudinal ADRC controller design} 
%%%%%%%%%%%%%%%%%%%%%%%%%%%%%%%%%%%%%%

In the same manner as in \ref{Section:Lateral ADRC design}, the along-track error dynamic can be represented in a more compact form:
\begin{equation}
    \dot{e}_\mathrm{s}(t) = -v(t) + f_\mathrm{v}(t),
    \label{eqn:Longitudinal_ADRC_control_channel}
\end{equation}
where $ f_\mathrm{v}(t)=v_\mathrm{L}(t)\cdot\cos{\theta_\mathrm{e}(t)}-v_\mathrm{d}(t)$ is a total disturbance in longitudinal control channel. 

The extended state-space form of \eqref{eqn:Longitudinal_ADRC_control_channel} is thus:
\begin{align}
    \begin{bmatrix} \dot{e}_\mathrm{s}(t)\\ \dot{f}_\mathrm{v}(t)\end{bmatrix} &=
    \underbrace{\begin{bmatrix} 0 & 1\\ 0 & 0\end{bmatrix}}_{\Matrix{A}_\mathrm{v}} \cdot \begin{bmatrix} e_\mathrm{s}(t)\\ f_\mathrm{v}(t)\end{bmatrix}+\underbrace{\begin{bmatrix} -1\\0\end{bmatrix}}_{\Matrix{B}_\mathrm{v}} \cdot v(t)+\begin{bmatrix} 0\\ 1\end{bmatrix}\cdot\dot{f}_\mathrm{v}(t).
    \label{eqn:Longitudinal_state_space_model}
\end{align}

In that case, the system states can be estimated by a longitudinal ESO having the following structure:
\begin{align}
    \begin{bmatrix} \hat{\dot{e}}_\mathrm{s}(t)\\  \hat{\dot{f}}_\mathrm{v}(t)\end{bmatrix}=\Matrix{A}_\mathrm{v} \cdot \begin{bmatrix} \hat{e}_{s}(t)\\  \hat{f}_\mathrm{v}(t)\end{bmatrix}+\Matrix{B}_\mathrm{v} \cdot v(t)+\begin{bmatrix} l_\mathrm{1v}\\l_\mathrm{2v}\end{bmatrix}\cdot\left[e_\mathrm{s}(t)-\hat{e}_{s}(t)\right],
    \label{eqn:Longitudinal_ESO}
\end{align}
where $l_\mathrm{1v}$, $l_\mathrm{2v}$ are the observer gains.  

The control signal for the longitudinal channel is thus:
\begin{equation}
   v(t)=-k_\mathrm{1v} \cdot \left[e_{\mathrm{s(ref)}}-\hat{e}_\mathrm{s}(t)\right]-\dot{e}_{\mathrm{s(ref)}}(t)+\hat{f}_\mathrm{v}(t), 
   \label{eqn:longitudinal_ADRC_law}
\end{equation}
where $k_\mathrm{1v}$ is controller adjusting parameter.

In the same manner as in the lateral ADRC controller tuning, the longitudinal controller and ESO gains are tuned using the \emph{bandwidth parameterization} \cite{1242516} methodology, i.e:
\begin{equation}
    \lambda + \omega_\mathrm{CL_v} \stackrel{!}{=} \lambda+
    k_\mathrm{1v},      
    \label{tuning_longitidunal}
\end{equation}
for the controller part and:
\begin{equation}
    \left( \lambda + \omega_\mathrm{ESO_v} \right)^2
    \stackrel{!}{=} \lambda^{2}  + l_\mathrm{1v}\cdot  \lambda + l_\mathrm{2v}, 
    \label{tuning_longitidunal_controller}
\end{equation}
for the observer part, where $\omega_\mathrm{CL_v}$ and $\omega_\mathrm{ESO_v}$ are the longitudinal closed-loop control system and longitudinal ESO bandwidths, respectively.

%%%%%%%%%%%%%%%%%%%%%%%%%%%%%%%%%%%%%%
%%%%%%%%%%%%%%%%%%%%%%%%%%%%%%%%%%%%%%
%%%%%%%%%%%%%%%%%%%%%%%%%%%%%%%%%%%%%%

\section{Simulation verification}
\label{sec:Section4}

%%%%%%%%%%%%%%%%%%%%%%%%%%%%%%%%%%%%%%
%%%%%%%%%%%%%%%%%%%%%%%%%%%%%%%%%%%%%%
%%%%%%%%%%%%%%%%%%%%%%%%%%%%%%%%%%%%%%

The simulations of UTV leader-follower control are carried out in MATLAB/Simulink using a model of a laboratory UTV Jaguar-Lite \cite{Jaguar_guide}, with parameters $r = 0.3\,$m and $B = 0.7\,$m. The proposed lateral and longitudinal ADRC controllers are compared with UTV control architecture realized by standard PID controller with noise filter in the lateral channel and PI controller in the longitudinal control channel. Furthermore, to ensure the simulations resemble real UTV constraints, the maximum angular velocity in both controllers (PI/PID and ADRC) is limited to $\dot{\theta}_\mathrm{max}=5\,$rad/s. The simulations are conducted in two leader-follower scenarios, described next.  

%%%%%%%%%%%%%%%%%%%%%%%%%%%%%%%%%%%%%%
\subsection{Simulation results (Scenario~\#1)}
%%%%%%%%%%%%%%%%%%%%%%%%%%%%%%%%%%%%%%

In this scenario, the leader trajectory is defined by the leader course angle: 
\begin{equation}
    \theta_\mathrm{L}(t)=
    \begin{cases}
    -0.12\cdot t, & \text{when} \quad  0\leq t<10,\\
    0.88\cdot t, & \text{when} \quad  10\leq t<15,\\
    1.07\cdot t, & \text{when} \quad  15\leq t<35,\\
    1.22\cdot t, &  \text{when} \quad  35\leq t<37,\\
    0.87\cdot t, &  \text{when} \quad  37 \leq t<40,\\
    1.17\cdot t, &  \text{when} \quad  40 \leq t<45,\\
    0.97\cdot t, &  \text{when} \quad  45\leq t<47,\\
    0.92\cdot t, &  \text{when} \quad  47\leq t<52,\\
    0.77\cdot t, &  \text{when} \quad  52\leq t<60.\\
    \end{cases} 
    \label{path_leader}
\end{equation}

In order to thoroughly analyze the performance of the designed UTV control under different working conditions, the simulation is divided into \Revised{five} intervals:
\begin{itemize}
    \item [i)] In the first interval (from 0 to 10 seconds), simulation is performed by assuming the perfect conditions without track slippage dynamics($a_\mathrm{L}=a_\mathrm{R}=$1) and measurement noises in the lateral and longitudinal channels, while the leader velocity and the reference along-track error are set as $v_\mathrm{L}(t)=2\,$m/s and ${e}_\mathrm{s(ref)}=2\,$m, respectively. It is assumed that UTV and leader have the same initial coordinates (i.e. $X=3\,$m and $Y=20\,$m).
    \item [ii)] In the second interval (from 10 to 15 seconds), the leader stops moving, i.e. $v_\mathrm{L}(t)=0\,$m/s.
    \item [iii)] In the third interval (from 15 to 30 seconds), the leader velocity is set as $v_\mathrm{L}(t)=2+1.4\sin{(t)}\,$m/s, and the complex slippage disturbances are assumed with track slippage coefficients $a_\mathrm{R}(t)=0.7+0.3\sin{(5t)}$ and $a_\mathrm{L}(t)=0.7+0.3\sin{(2t)}$.
    \item [iv)] In the fourth interval (from 30 to 45 seconds), the previous conditions were kept, and in addition, measurement noises, modeled as Gaussian noises with standard deviations $0.02\,$m in the lateral channel and $0.01\,$m in the longitudinal channel, are included.
    \Revised {\item [v)] In the fifth and last interval (from 45 to 60 seconds), in order to analyze the effect of changing the defined reference along-track error, it is increased to $e_\mathrm{s(ref)}=3\,$m, while other simulation parameters are kept as in the previous interval.}
\end{itemize}

To enable a fair comparison between the two considered control algorithms, they are empirically tuned to have similar tracking performance in the absence of disturbances (first interval). This approach ensures that both structures are initially set up to achieve comparable tracking capabilities under ideal conditions. Accordingly, for the lateral PID controller, the proportional, integral, derivative gain and the noise filter coefficients are adopted as $K_\mathrm{pl}=4$, $K_\mathrm{il}=2$, $K_\mathrm{dl}=0.5$, and $N=50$, respectively, while the proportional and integral gains of the longitudinal PI controller are set as $K_\mathrm{pv}=3$, $K_\mathrm{iv}=3$. The parameters for the lateral and longitudinal ADRC controllers are set based on the proposed \emph{bandwidth parametrization} method by choosing $\omega_\mathrm{CL_l}=1.2\,$rad/s and $\omega_\mathrm{ESO_l}=10\,$rad/s, $\omega_\mathrm{CL_v}=1\,$rad/s and $\omega_\mathrm{ESO_v}=10\,$rad/s, while the parameter $b_0$ in lateral ADRC controller is set according to the value of the leader longitudinal velocity in the first interval, i.e. $b_0=-2$. The tuned controller parameters values are kept for the remainder of the simulation time. 

\begin{figure}
     \centering
     \begin{subfigure}[b]{\textwidth}
         \centering
         \includegraphics[width=0.75\textwidth]{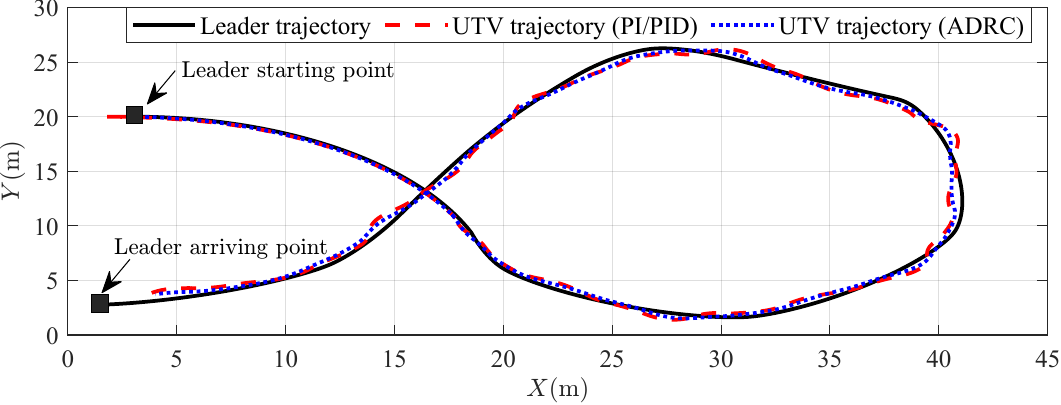}
         \caption{Leader and UTV trajectories}
         \label{fig:path}
     \end{subfigure}
     \hfill
     \begin{subfigure}[b]{\textwidth}
         \centering
         \includegraphics[width=0.75\textwidth]{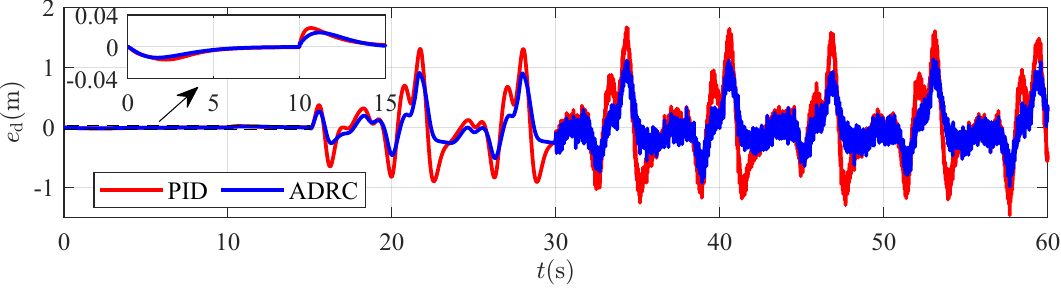}
         \caption{Cross-track error}
         \label{fig:ed_3senario}
     \end{subfigure}
     \hfill
     \begin{subfigure}[b]{\textwidth}
         \centering
         \includegraphics[width=0.75\textwidth]{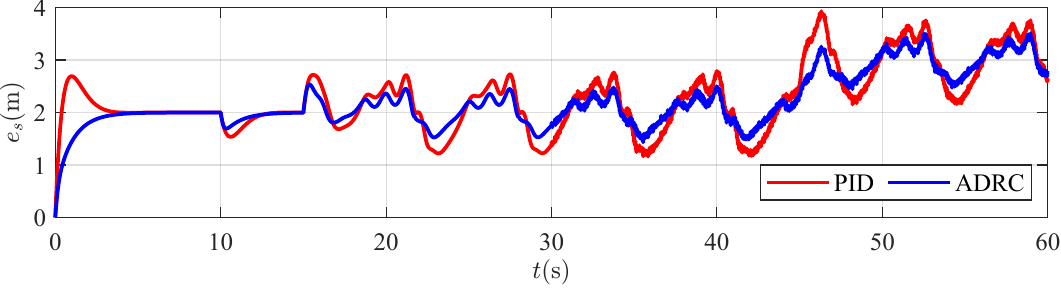}
         \caption{Along-track error}
         \label{fig:es_3senario}
     \end{subfigure}
          \hfill
     \begin{subfigure}[b]{\textwidth}
         \centering
         \includegraphics[width=0.75\textwidth]{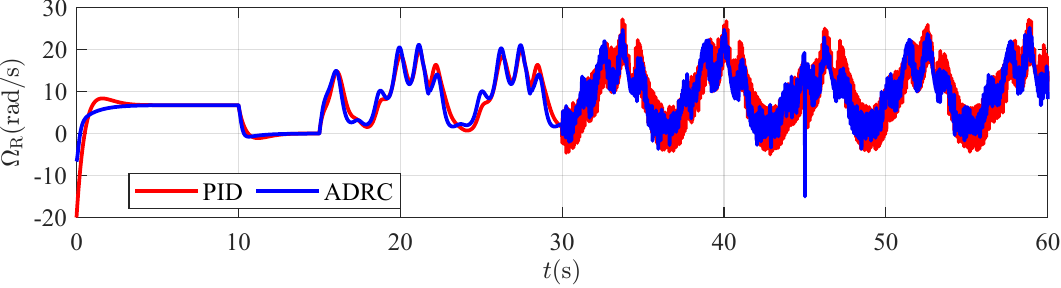}
         \caption{Right wheel angular velocity}
         \label{fig:Right wheel_3senario}
     \end{subfigure}
          \hfill
     \begin{subfigure}[b]{\textwidth}
         \centering
         \includegraphics[width=0.75\textwidth]{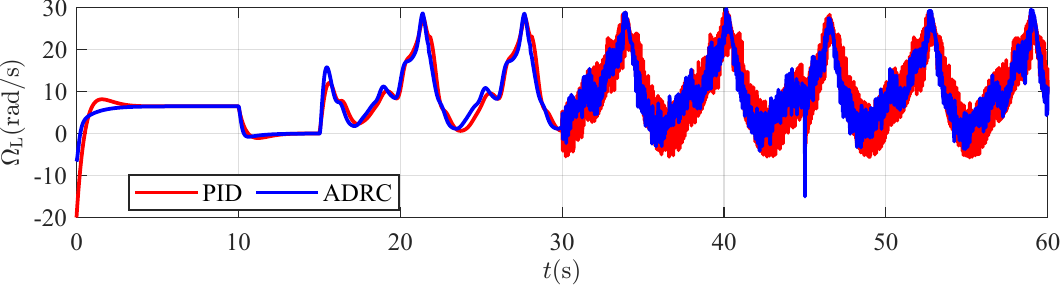}
         \caption{Left wheel angular velocity}
         \label{fig:Left wheel_3senario}
     \end{subfigure}
        \caption{Simulation results (for Scenario~\#1).}
        \label{fig:simulation_results}
\end{figure}

The simulation results, which include leader trajectory tracking, cross-track errors, along-track errors, and track wheel velocities are gathered in \refFig{fig:simulation_results}. \Revised{To quantitatively analyze the obtained simulation results, integrals of the absolute cross-track errors ($\lvert e_\mathrm{d}(t) \rvert$) and the difference between reference and obtained along-track errors ($e_\mathrm{s(e)}=\lvert e_\mathrm{s(ref)}(t)-e_\mathrm{s}(t) \rvert$) during each simulation intervals are calculated and compared in Table~\ref{tab:features} for tested control systems with PID and ADRC controllers.}

\begin{table*}[htp]
\centering
\caption{\Revised{Integral of the absolute cross-track error and difference between reference and obtained along-track errors for systems with PID and ADRC controllers in the simulation Scenario~\#1.}}

\label{tab:features}

\resizebox{\textwidth}{!}{%

\begin{tabular}{lcccccccccccccc}
\toprule
& \multicolumn{2}{c}{\multirow{2}{*}{First interval} \vspace{5pt}} & \multicolumn{2}{c}{\multirow{2}{*}{Second interval}} & \multicolumn{2}{c}{\multirow{2}{*}{Third interval}} & \multicolumn{2}{c}{\multirow{2}{*}{Fourth interval}} & \multicolumn{2}{c}{\multirow{2}{*}{Fifth interval}} \\

& \multicolumn{2}{c}{(from 0 to 10$\,$s)} & \multicolumn{2}{c}{(from 10 to 15$\,$s)} & \multicolumn{2}{c}{(from 15 to 30$\,$s)} & \multicolumn{2}{c}{(from 30 to 45$\,$s)} & \multicolumn{2}{c}{(from 45 to 60$\,$s)} \\

\midrule

{} &  $e_\mathrm{d}$ &  $e_\mathrm{s(e)}$ &  $e_\mathrm{d}$ &  $e_\mathrm{s(e)}$ &  $e_\mathrm{d}$ &  $e_\mathrm{s(e)}$ &  $e_\mathrm{d}$ &  $e_\mathrm{s(e)}$ &  $e_\mathrm{d}$ &  $e_\mathrm{s(e)}$ \\

\midrule

PID & 0.059 & 1.318 & 0.056 & 0.667 & 5.891 & 6.183 & 6.329 & 6.188 & 7.052 & 6.385 \\

ADRC & 0.050 & 1.583 & 0.050 & 0.414 & 3.439 & 3.483 & 3.720 & 3.255 & 4.220 & 3.654 \\

\bottomrule

\end{tabular}

}
\end{table*}

Based on the obtained simulation results, it is evident that during the first interval, without the slippage disturbance effect, both control approaches effectively track the leader's trajectory and exhibit similar performance. The velocities of the drive wheels are smoothly controlled in both systems, although the PID approach produces higher peak values at the start of the simulation.

In the second interval, when the leader stops moving, it is noteworthy that both systems are able to maintain stable tracking, without oscillating in the along-track error dynamics. However, one can see that the ADRC approach demonstrates a better transient response compared to the PI/PID approach.

The third interval is crucial for assessing the robustness of the control algorithms against slippage disturbances. The results demonstrate the effectiveness of the ADRC-based approach in handling slippage disturbances. \Revised {In fact, the ADRC structure achieves a significantly lower value of the integrals of absolute cross-track and along-track errors} as well as lower peaks of both errors, compared to the PI/PID structure and given similar values of the drive wheel velocities.

Observing the simulation results in the fourth interval, it can be noted that ADRC outperforms the PI/PID approach under measurement noise conditions. The ADRC maintains accurate leader tracking and generates smoother drive wheel velocities compared to PI/PID, further validating its effectiveness for practical implementation.

\Revised{Finally, from the simulation results in the last, fifth interval, it is evident that the increase of the reference along-track error had a negligible impact on the tracking performance degradation and that the ADRC structure retained its advantages over the PID structure in this test.}

%%%%%%%%%%%%%%%%%%%%%%%%%%%%%%%%%%%%%%
\Revised{\subsection{Simulation results (Scenario~\#2)}}
%%%%%%%%%%%%%%%%%%%%%%%%%%%%%%%%%%%%%%

\begin{figure}[htb!]
    \centering
    \includegraphics[width=0.75\columnwidth]{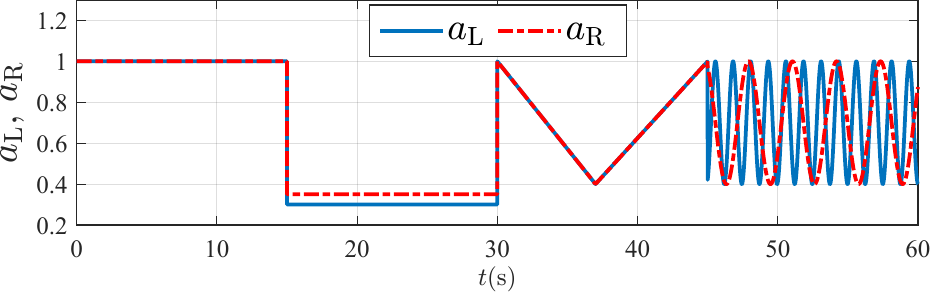} 
     \caption{\Revised{User-defined slippage disturbance in simulation Scenario~\#2.}}    
     \label{fig:aRL}
\end{figure}

\Revised{In this simulation scenario, the performance of the considered leader-follower control algorithms in the presence of slippage disturbances with various dynamics is analyzed. The controllers' settings are kept as in the previous scenario, while the leader trajectory is defined as a circular path this time, with fixed leader velocity ${v}_\mathrm{L}=2\,$m/s and reference along-track error ${e}_\mathrm{s(ref)}=2\,$m. It is assumed that the UTV and the leader have the same initial coordinates (i.e. $X=25$m and $Y=45\,$m). The simulation is divided into the following four intervals with different user-defined dynamics of the friction coefficients (i.e. the slippage disturbances), visualized in Figure~\ref{fig:aRL}:} 
\begin{itemize}
    \item [i)] \Revised{In the first interval (from 0 to 15 seconds), simulation is performed without track slippage dynamics ($a_\mathrm{L}=a_\mathrm{R}=$1).} 
    \item [ii)] \Revised{In the second interval (from 15 to 30 seconds), constant slippage disturbances for the left and right tracks are assumed.} 
    \item [iv)] \Revised{In the third interval (from 30 to 45 seconds), ramp slippage disturbances are assumed.} 
    \item [iv)] \Revised{In the fourth interval (from 45 to 60 seconds), sinusoidal slippage disturbances are assumed.}    
\end{itemize}

\begin{figure}
     \centering
     \begin{subfigure}[b]{\textwidth}
         \centering
         \includegraphics[width=0.75\textwidth]{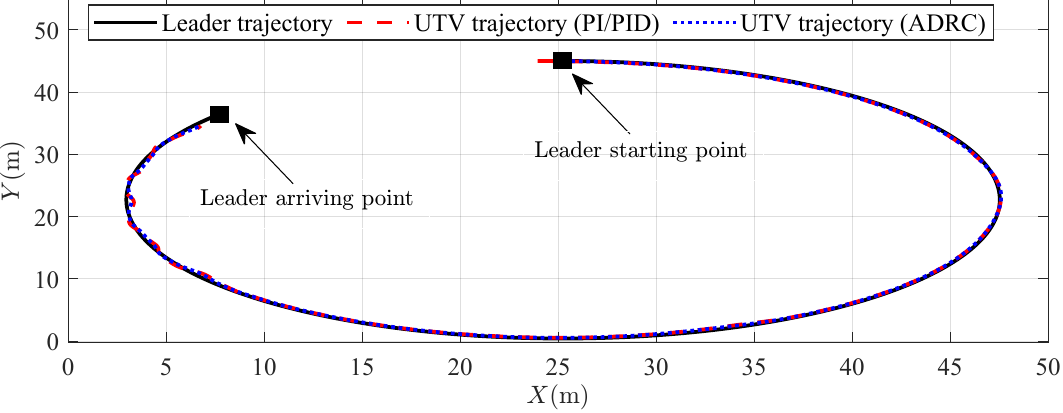}
         \caption{Leader and UTV trajectories}
         \label{fig:path_disturbance}
     \end{subfigure}
     \hfill
     \begin{subfigure}[b]{\textwidth}
         \centering
         \includegraphics[width=0.75\textwidth]{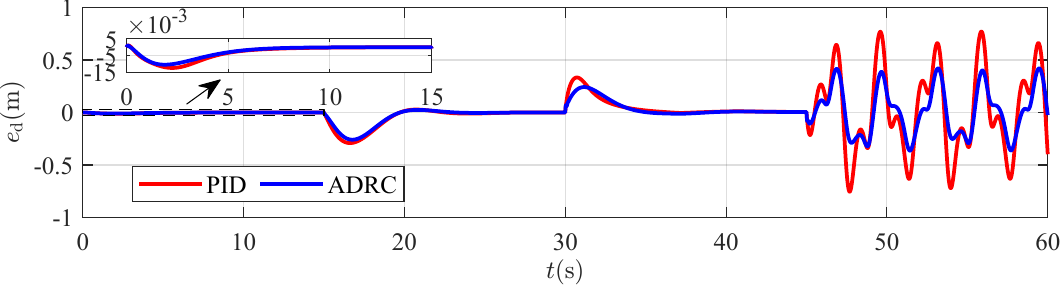}
         \caption{Cross-track error}
         \label{fig:ed_disturbance}
     \end{subfigure}
     \hfill
     \begin{subfigure}[b]{\textwidth}
         \centering
         \includegraphics[width=0.75\textwidth]{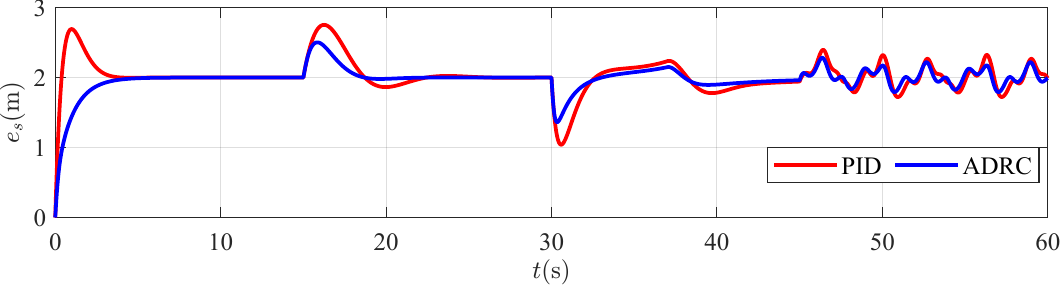}
         \caption{Along-track error}
         \label{fig:es_disturbance}
     \end{subfigure}
          \hfill
     \begin{subfigure}[b]{\textwidth}
         \centering
         \includegraphics[width=0.75\textwidth]{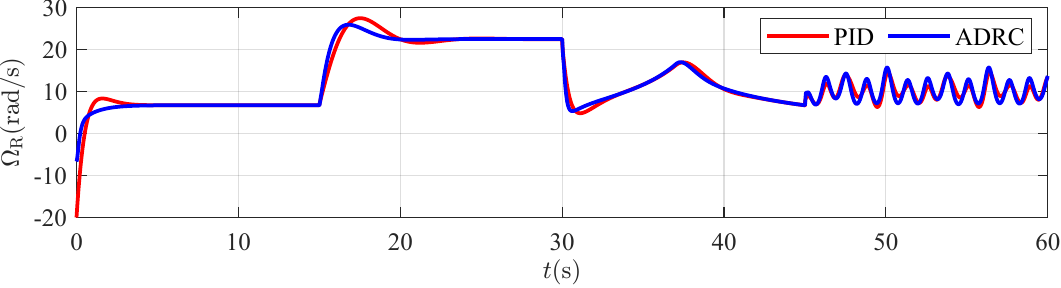}
         \caption{Right wheel angular velocity}
         \label{fig:Right wheel_disturbance}
     \end{subfigure}
          \hfill
     \begin{subfigure}[b]{\textwidth}
         \centering
         \includegraphics[width=0.75\textwidth]{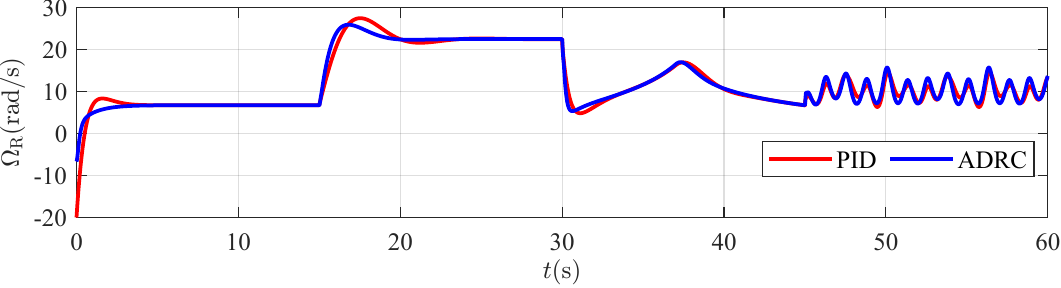}
         \caption{Left wheel angular velocity}
         \label{fig:Left wheel_disturbance}
     \end{subfigure}
        \caption{Simulation results (for Scenario~\#2).}
        \label{fig:simulation_results_disturbance}
\end{figure}

\Revised{ 
The simulation results, which include leader trajectory tracking, cross-track errors, along-track errors, and track wheel velocities are gathered in \refFig{fig:simulation_results_disturbance}. In the same way as in the previous scenario, the integrals of the absolute cross-track error and the difference between the reference and obtained along-track errors ($e_\mathrm{s(e)}$) are computed for each simulation interval and compared in Table~\ref{tab:IAE_disturbance} for control systems with PID and ADRC.}

\begin{table*}[htp]
\centering
\caption{\Revised{Integral of the absolute cross-track error and difference between reference and obtained along-track errors for systems with PID and ADRC controllers in the simulation Scenario~\#2.}}

\label{tab:IAE_disturbance}

\resizebox{\textwidth}{!}{%

\begin{tabular}{lcccccccccccc}
\toprule
& \multicolumn{2}{c}{\multirow{2}{*}{First interval} \vspace{5pt}} & \multicolumn{2}{c}{\multirow{2}{*}{Second interval}} & \multicolumn{2}{c}{\multirow{2}{*}{Third interval}} & \multicolumn{2}{c}{\multirow{2}{*}{Fourth interval}}  \\

& \multicolumn{2}{c}{(from 0 to 15$\,$s)} & \multicolumn{2}{c}{(from 15 to 30$\,$s)} & \multicolumn{2}{c}{(from 30 to 45$\,$s)} & \multicolumn{2}{c}{(from 45 to 60$\,$s)}  \\

\midrule

{} &  $e_\mathrm{d}$ &  $e_\mathrm{s(e)}$ &  $e_\mathrm{d}$ &  $e_\mathrm{s(e)}$ &  $e_\mathrm{d}$ &  $e_\mathrm{s(e)}$ &  $e_\mathrm{d}$ &  $e_\mathrm{s(e)}$ &  \\

\midrule

PID &     0.0451	&   11.3195	  &  0.8716	  &  2.0132	 &   0.7671	 &   2.8979	  &  4.9933	 &   2.0054 \\

ADRC &     0.0377	&   11.5815 &	    0.7485	&    0.9617	&    0.6920	   & 1.6281	  &  2.8620	 &   1.3692 \\

\bottomrule

\end{tabular}

}

\end{table*}

\Revised{Based on the obtained results, it is evident that, compared to PI/PID, the ADRC algorithm achieves higher tracking performance in the presence of all analyzed types of slippage disturbances. Actually, in the case of step slippage disturbances, both algorithms manage to minimize tracking errors but the transient performances are better for ADRC, especially in the tracking of the defined reference along-track error. A similar observation can be made when dealing with ramp slippage disturbances (third interval), where the calculated integrals of absolute values of errors are significantly smaller for ADRC. The benefits of the ADRC algorithm were confirmed in the fourth interval as well, where ADRC achieves lower peak values and the integrals of the absolute tracking errors.}

%%%%%%%%%%%%%%%%%%%%%%%%%%%%%%%%%%%%%%
%%%%%%%%%%%%%%%%%%%%%%%%%%%%%%%%%%%%%%
%%%%%%%%%%%%%%%%%%%%%%%%%%%%%%%%%%%%%%

\section{Experimental validation}
\label{sec:Section5}

%%%%%%%%%%%%%%%%%%%%%%%%%%%%%%%%%%%%%%
%%%%%%%%%%%%%%%%%%%%%%%%%%%%%%%%%%%%%%
%%%%%%%%%%%%%%%%%%%%%%%%%%%%%%%%%%%%%%

The experimental validation of the proposed leader-follower ADRC algorithm for UTV is realized by the Jaguar-Lite UTV \cite{Jaguar_guide}, equipped with a camera, laser sensor, and WiFi modules for sending sensor data and receiving UTV control signals. \Revised{The camera is mounted on the front of the vehicle at a height of $0.1\,$m above the ground.} The experiments are carried out with a human leader moving in front of the vehicle. The detailed experimental setup and obtained results are described below.  

%%%%%%%%%%%%%%%%%%%%%%%%%%%%%%%%%%%%%%
\subsection{Test setup and technical realization}
%%%%%%%%%%%%%%%%%%%%%%%%%%%%%%%%%%%%%%

\begin{figure}[htb!]
    \centering
    \includegraphics[width=.9\columnwidth]{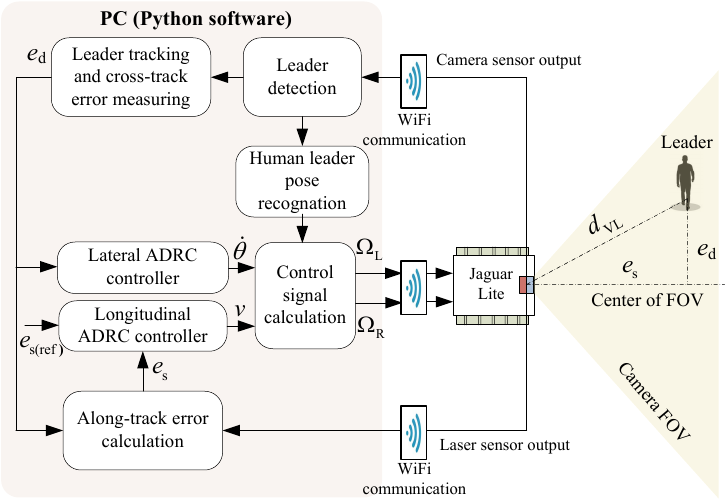} 
     \caption{Block diagram of the prepared experimental setup.}    
     \label{fig:Proposed_follow_me_algorithm}
\end{figure}

Based on the designed UTV leader-follower control algorithm, it is evident that the knowledge of the current cross-track and along-track errors is essential for practical implementation. These errors will serve as feedback signals for the controllers' design in the lateral and longitudinal channels. Therefore, a camera-based algorithm for the cross-track error measuring, and a laser-based system for the along-track error calculation are proposed. In addition, to further enhance the capabilities of the proposed framework, the camera sensor is utilized for leader pose recognition, specifically detecting a human leading the vehicle. This feature enables the UTV to respond to the leader's commands, such as stopping or accelerating towards or away from the leader, in specific situations. By leveraging the camera-based leader pose recognition, the system achieves improved responsiveness and adaptability, facilitating effective human-vehicle interaction. The block diagram of the experimental setup is shown in \refFig{fig:Proposed_follow_me_algorithm}.% The proposed blocks are implemented in Python software and their structures are shown in follows.  

%%%%%%%%%%%%%%%%%%%%%%%%%%%%%%%%%%%%%%
\subsubsection{Leader detection, tracking, and leader-follower error measuring}
%%%%%%%%%%%%%%%%%%%%%%%%%%%%%%%%%%%%%%

Both detection and tracking of the human leader are achieved by analyzing frames of the video stream captured by the UTV camera, using the commercial YOLOv8 \cite{original_YOLO,ComparaisonYOLO8} algorithm in conjunction with the predefined \textit{person} class \cite{ultralytics2021yolov8docs}.

%To simplify the case, we assume the presence of a single leader that the UTV is following in the scene.

\begin{figure}[htb!]
    \centering
    \includegraphics[width=0.45\columnwidth]{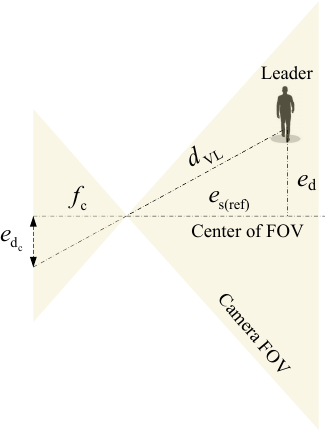} 
     \caption{\Revised{Diagram for calculating cross-track error (SI units: meters).}}     
     \label{fig:Calculation_e_d}
\end{figure}

To calculate the cross-track error for each video frame, defined in the discrete-time $kT_\mathrm{s}$ (where $T_\mathrm{s}$ is the sample time), the bounding box of the detected and tracked human leader is obtained and denoted as $D_\mathrm{L}(kT_\mathrm{s})$. Then, the horizontal coordinate of the center of $D_\mathrm{L}(kT_\mathrm{s})$ is calculated as $C_\mathrm{{D_\mathrm{L}}}(kT_\mathrm{s})$, and the cross-track error in pixel $e_\mathrm{d_{c}}(kT_\mathrm{s})$ is defined as the horizontal distance between $C_{\mathrm{D}_\mathrm{L}}(kT_\mathrm{s})$ and the horizontal pixel coordinate of the camera center field of view (FOV). To enable $e_\mathrm{d}(kT_\mathrm{s})$ in meters, the transformation based on the assumed along-track error $e_\mathrm{s(ref)}$ is used. \Revised{The appropriate diagram for calculation of $e_\mathrm{d}(kT_\mathrm{s})$ is shown in \refFig{fig:Calculation_e_d}, where one can see that:}
\begin{equation}
    \frac{e_\mathrm{d}(kT_\mathrm{s})}{e_\mathrm{s(ref)}} =\frac{e_\mathrm{d_\mathrm{c}}(kT_\mathrm{s})\cdot P}
    {f_\mathrm{c}},  
    \label{eqn:calculation_ed_metre}
\end{equation}
with:
\begin{equation}
    e_\mathrm{d}(kT_\mathrm{s}) = \frac{e_\mathrm{s(ref)} \cdot e_\mathrm{d_\mathrm{c}}(kT_\mathrm{s}) }{ f_\mathrm{c}}  \cdot P,
    \label{eqn:ed_metre}
\end{equation}
where $f_\mathrm{c}$  denotes the camera focal length and $P$ represents the pixel pitch dimension.

In this experiment, a camera with the resolution of 1280 $\times$ 720 pixels is used, with a focal length of $f_\mathrm{c} = 2.8\,$mm and the pixel pitch dimension $P = 19 \, \mu \mathrm{m}$. 

Calculating the along-track leader follow error is realized based on the previously calculated value $e_\mathrm{d}(kT_\mathrm{s})$ as:
\begin{equation}
        e_\mathrm{s}(kT_\mathrm{s}) = \sqrt{d_\mathrm{VL}^2(kT_\mathrm{s}) - e_\mathrm{d}^2(kT_\mathrm{s})},  
        \label{eqn:laser_based_alog_track_error_measuring}
\end{equation}
where $d_\mathrm{VL}(kT_\mathrm{s})$ represents the laser-based measured distance between the leader and the UTV (see \refFig{fig:Proposed_follow_me_algorithm}). 

%%%%%%%%%%%%%%%%%%%%%%%%%%%%%%%%%%%%%%
\subsubsection{Human leader pose recognition}
%%%%%%%%%%%%%%%%%%%%%%%%%%%%%%%%%%%%%%

\begin{figure}[ht] 
 \centering
 \includegraphics[width=\columnwidth]{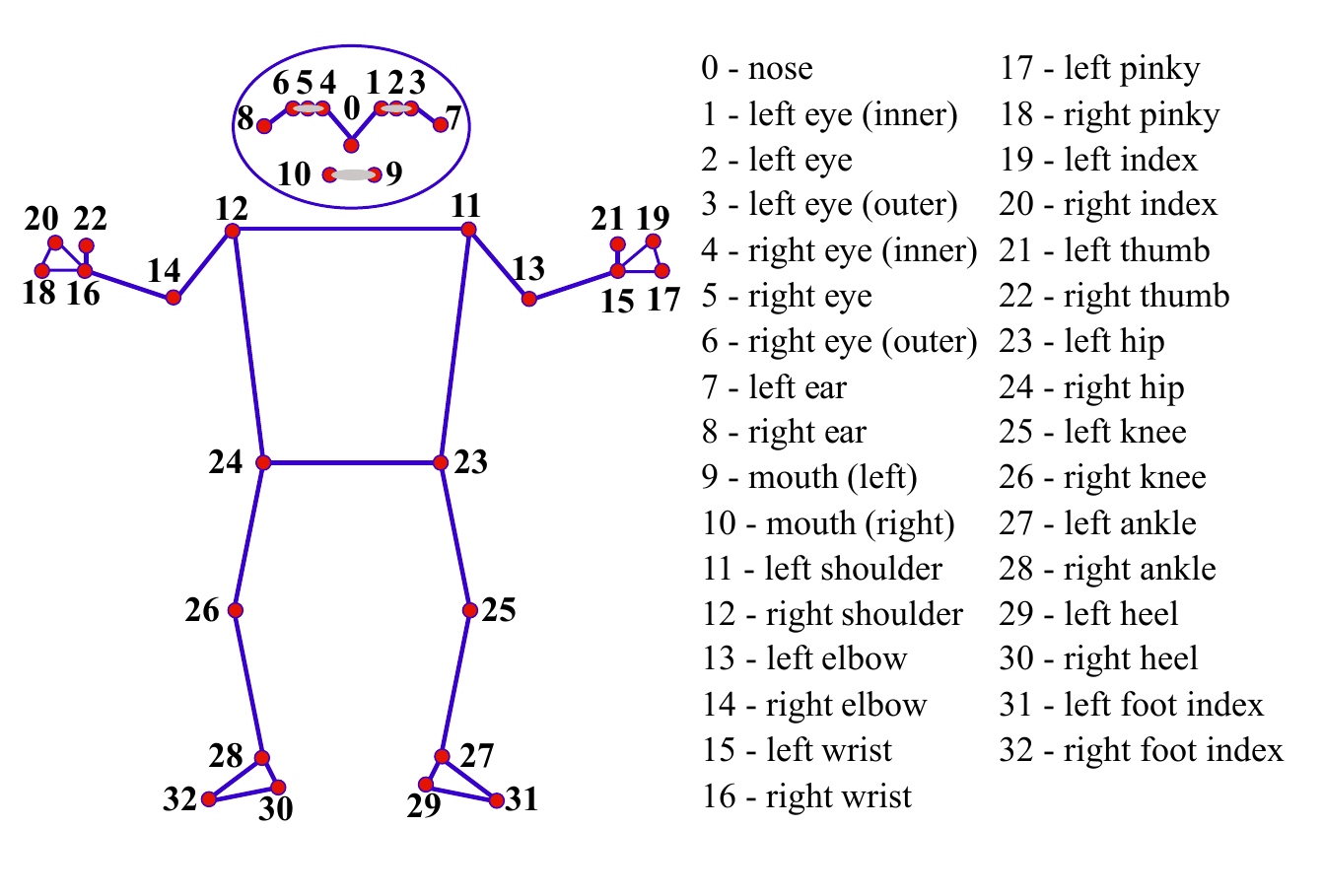} 
 \caption{Definition of landmarks in Media Pipe Pose algorithm \cite{mediapipe}.} 
 \label{fig:landmarks}
\end{figure}

The human leader pose recognition is realized by Media Pipe Pose (MPP) algorithm  \cite{mediapipe} using the extracted 2D landmarks on the human body \cite{blog-google-mediapipe}, shown in \refFig{fig:landmarks}. However, in this case, a specific set of 20 landmarks (11, 12, 13, 14, 15, 16, 17, 18, 19, 20, 21, 22, 23, 24, 25, 26, 27, 28, 29, and 30.) is selected from the overall set of estimated landmarks and used to recognize four human leader poses: upright leader pose, crouching leader pose, crouching leader pose with right hand raised, and crouching leader pose with left hand raised. 

%%%%%%%%%%%%%%%%%%%%%%%%%%%%%%%%%%%%%%
\subsubsection{ADRC software implementation}
%%%%%%%%%%%%%%%%%%%%%%%%%%%%%%%%%%%%%%

\begin{figure}
     \centering
     \begin{subfigure}[b]{\textwidth}
         \centering
         \includegraphics[width=0.65\textwidth]{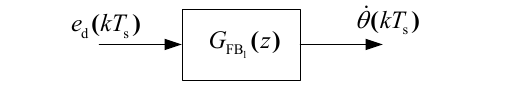}
         \caption{Discrete-time transfer function representation of lateral ADRC controller}
         \label{fig:Discrete_lateral}
     \end{subfigure}
     \hfill
     \begin{subfigure}[b]{\textwidth}
         \centering
         \includegraphics[width=0.65\textwidth]{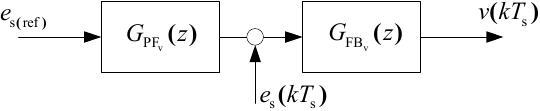}
         \caption{Discrete-time transfer function representation of longitudinal ADRC controller}
         \label{fig:Discrete_longitudinal}
     \end{subfigure}
        \caption{Discrete-time transfer function representations of (a) lateral ADRC controller and (b) longitudinal ADRC controller.}
        \label{fig:Discrete_ADRC}
\end{figure}

The designed ADRC lateral and longitudinal controllers \eqref{eqn:Lateral_ADRC_control_law} and \eqref{eqn:longitudinal_ADRC_law} are implemented in Python software based on their discrete-time transfer function representation \cite{herbst2023tuning}, and the obtained structures are shown in \refFig{fig:Discrete_ADRC}. Due to this paper's space limitation, the comprehensive procedure of deriving the discrete-time versions of the controllers is omitted here, and the obtained controller discrete-time transfer functions for both lateral and longitudinal channels are given in the Appendix. \Revised{The tuning parameters of both lateral and longitudinal ADRC controllers are set as in Section~\ref{sec:Section4}. The reference along-track error is set to $e_\mathrm{s(ref)}=2\,$m to allow the UTV camera to relatively accurately perceive and recognize the leader poses}.

%%%%%%%%%%%%%%%%%%%%%%%%%%%%%%%%%%%%%%
\subsubsection{Control signal calculation}
%%%%%%%%%%%%%%%%%%%%%%%%%%%%%%%%%%%%%%

\begin{figure}[htb!]
    \centering
    \includegraphics[width=.85\columnwidth]{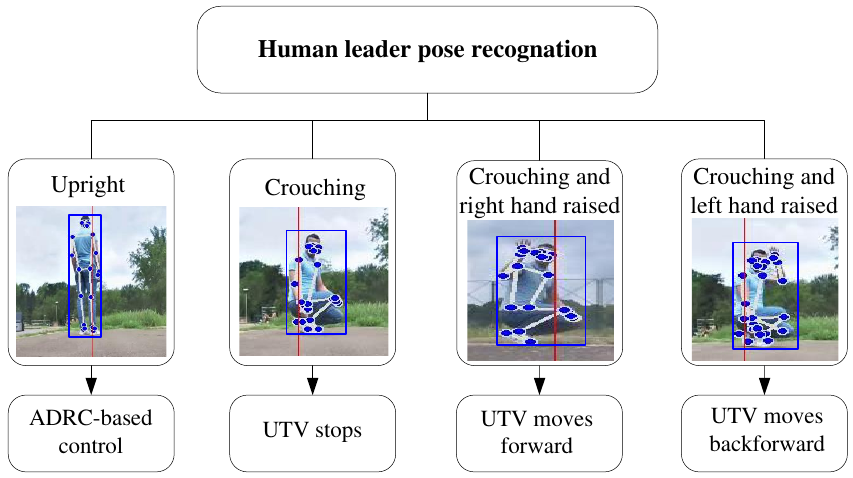} 
     \caption{Graphical representation of the control signal calculation block.}    
     \label{fig:control signal calculations diagram}
\end{figure}

The control signal calculation block aims to obtain the appropriate UTV control signals $\Omega_\mathrm{R}(t)$ and $\Omega_\mathrm{L}(t)$ using the pose recognition block output and the lateral and longitudinal controllers signals $\dot\theta(t)$ and $v(t)$. The graphical representation of the used algorithm is shown in \refFig{fig:control signal calculations diagram}, where one can see that in case when human pose is upright, the calculation of the control commands is realized by the lateral and longitudinal ADRC controller outputs, which are transformed to the appropriate UTV control signals $\Omega_\mathrm{R}(t)$ and $\Omega_\mathrm{L}(t)$ based on \eqref{eqn:LongitudinalAngular_vehicle_velocity_ideal}. On the other hand, when the leader pose is crouching, the UTV stops, while for the leader crouching poses with raised right and left hand, the UTV moves forward and backward, respectively, with constant longitudinal velocity and zero angular velocity.  
 
%%%%%%%%%%%%%%%%%%%%%%%%%%%%%%%%%%%%%%
\subsection{Experimental results}
%%%%%%%%%%%%%%%%%%%%%%%%%%%%%%%%%%%%%%

\begin{figure*}[]
\centering
    \subcaptionbox{Leader and UTV trajectories\label{fig:path_exp}}{
        \includegraphics[width=0.75\textwidth]{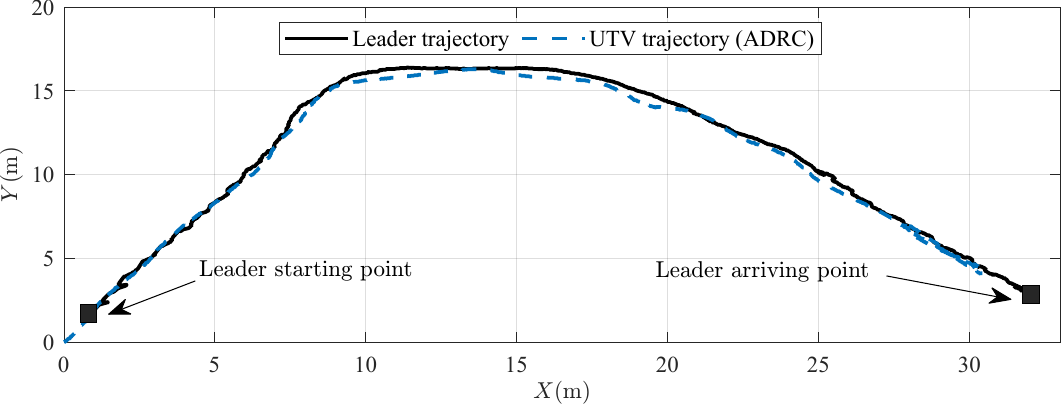}
    }\vspace{1mm} % Adjust the vertical spacing between subfigures

    \subcaptionbox{Cross-track error\label{fig:ed_exp}}{
        \includegraphics[width=0.75\textwidth]{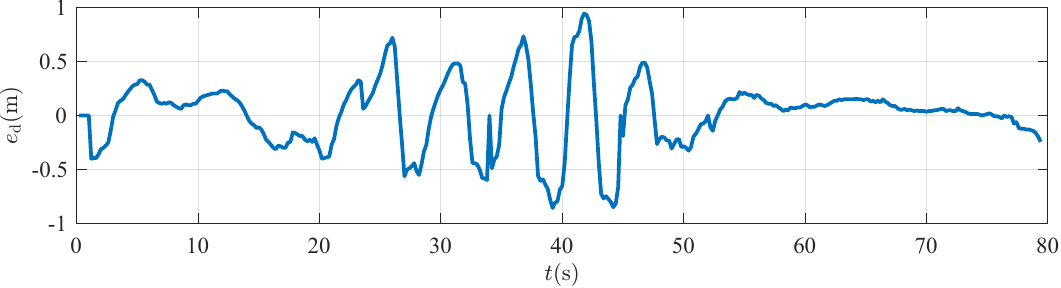}
    }\vspace{1mm} % Adjust the vertical spacing between subfigures
    
    \subcaptionbox{Along-track error\label{fig:es_exp}}{
        \includegraphics[width=0.75\textwidth]{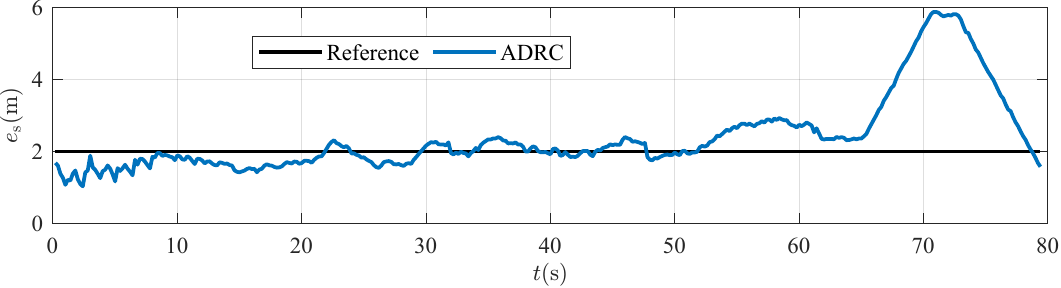}
    }\vspace{1mm} % Adjust the vertical spacing between subfigures

    \subcaptionbox{Right wheel angular velocity\label{fig:Right wheel_exp}}{
        \includegraphics[width=0.75\textwidth]{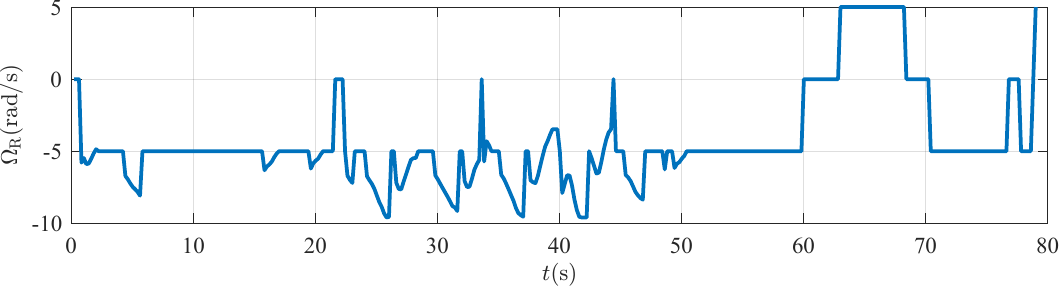}
    }\vspace{1mm} % Adjust the vertical spacing between subfigures
    
    \subcaptionbox{Left wheel angular velocity\label{fig:Left wheel_exp}}{
        \includegraphics[width=0.75\textwidth]{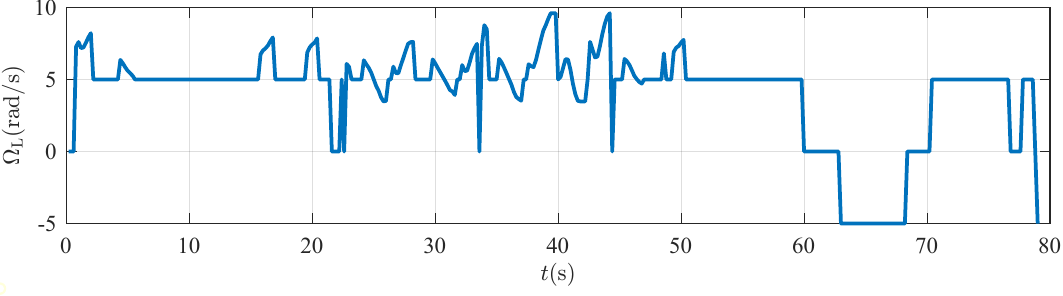}
    }
    \caption{Experimental results.}
    \label{fig:Experimental_results}
\end{figure*}

\begin{figure}[th!]
\centering
    \begin{subfigure}{\columnwidth}
    \centering
      \includegraphics[width=.65\columnwidth]{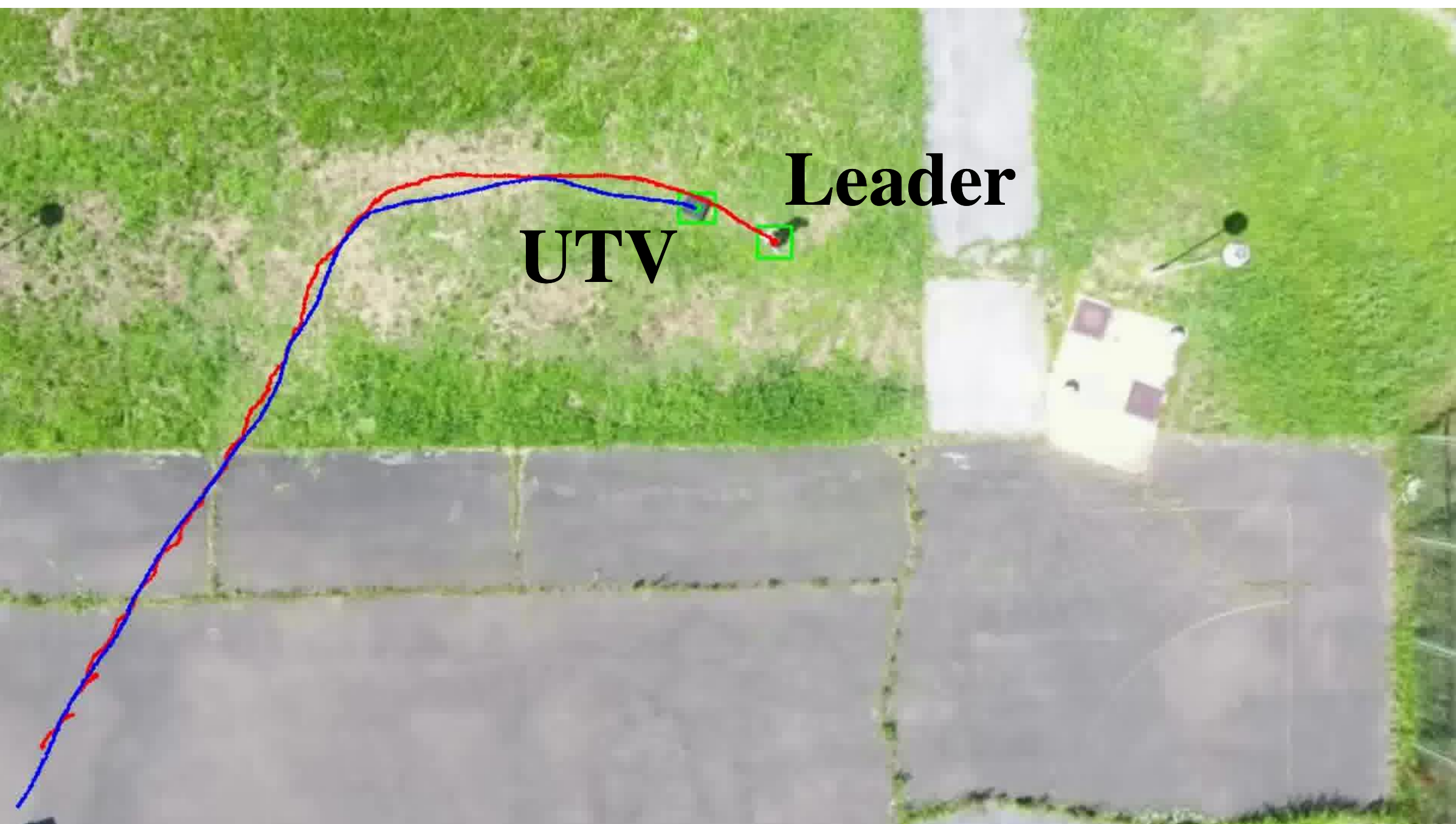}
      \caption{\Revised{Bird's-eye view (captured with a drone)}}
    \end{subfigure}
    \begin{subfigure}{\columnwidth}
    \centering
      \includegraphics[width=.65\columnwidth]{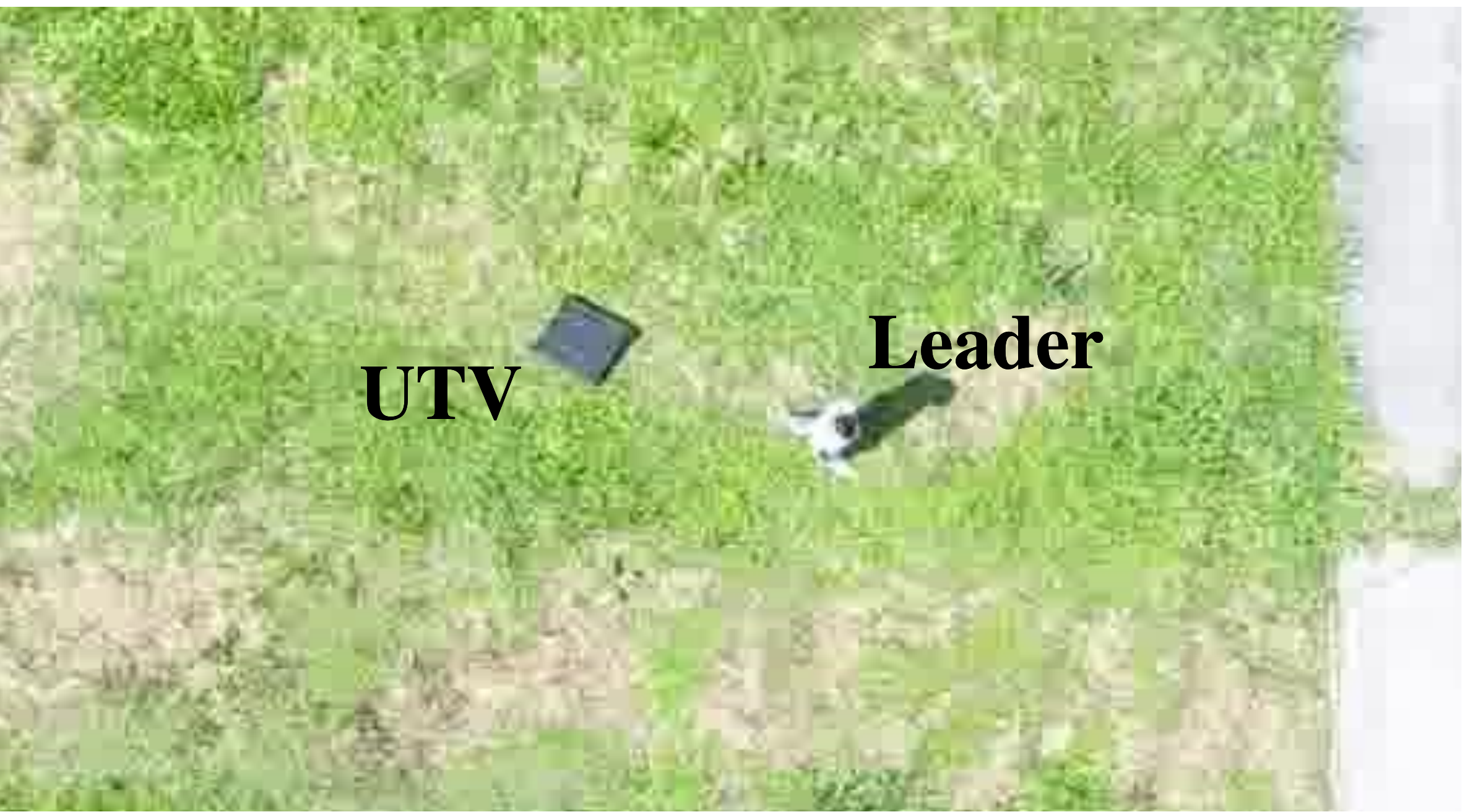}
      \caption{\Revised{Bird's-eye view (captured with a drone; zoomed in)}}
    \end{subfigure}
    \begin{subfigure}{\columnwidth}
    \centering
      \includegraphics[width=.65\columnwidth]{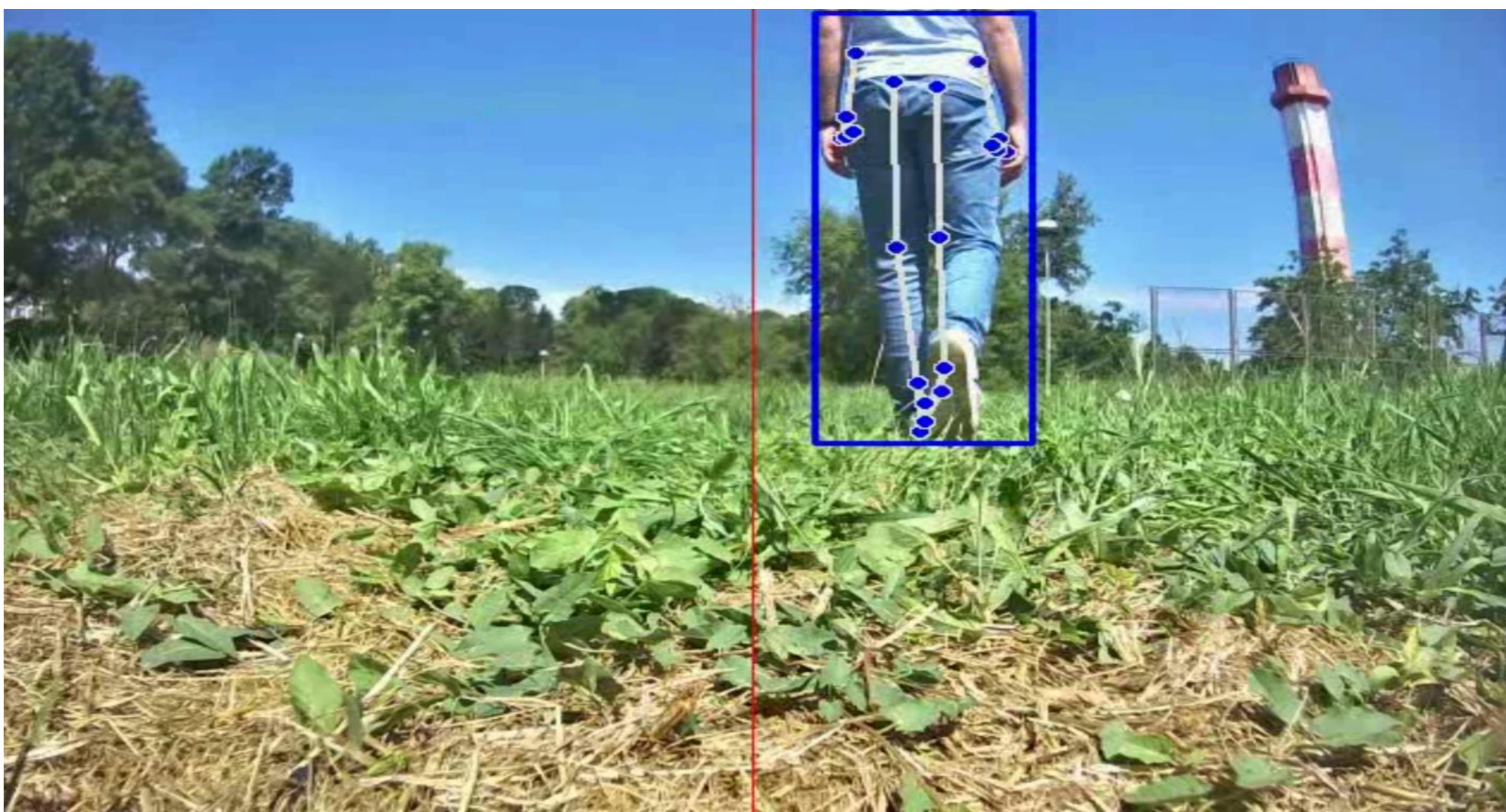}
      \caption{\Revised{UTV on-board camera view}}
    \end{subfigure}
  \caption{\Revised{Different views on the performed field tests.}}
  \label{fig:Expirement_test}
\end{figure}

To evaluate the performance of the developed Jaguar-Lite UTV leader-follower control algorithm, experiments with a human leader on the dirt-asphalt training ground are carried out. The longitudinal and lateral ADRC controllers are tuned as in \refSec{sec:Section4}, and discretization is performed with the sample time $T_\mathrm{s}=0.2\,$s, which corresponds with the predefined vehicle WiFi module sending rate. During the experiments, the human leader moved in front of the UTV, assuming various positions and issuing commands for the UTV movement. In the final interval (between 65 and 80 seconds), the leader assumes a crouched position to generate a stop signal for UTV, and then the leader raises the left and right hands as a command for the UTV to start moving backward and towards the leader, respectively. The experimental setup, as seen from a bird's-eye view and with a camera mounted on the Jaguar-Lite UTV, is depicted in \refFig{fig:Expirement_test}. The corresponding video of the conducted experimental test is given as the attached material of this paper (see a link to the corresponding YouTube video). Furthermore, the graphical representation of the obtained results, which include leader and UTV trajectories, the corresponding cross-track and along-track errors, and UTV velocity of the drive wheels, are shown in \refFig{fig:Experimental_results}. 

From the obtained results one can see that the UTV demonstrates robust detection and tracking capabilities of the leader's movement in both asphalt and rough dirt terrains, minimizing the cross-track error and maintaining the along-track error to the desired distance. \Revised {As expected, the errors were increased when the UTV moved over to the dirt terrain (approximately from 20 to 50 seconds) due to the significantly higher dynamics of the track slippage on such terrain. Nevertheless, one can see that the error peaks are close to those obtained through simulation analysis.}

The right and left wheel velocities exhibit synchronized patterns, indicating balanced movement of the UTV throughout the experiment as shown in Figures~\ref{fig:Right wheel_exp} and~\ref{fig:Left wheel_exp}. Slight variations in the wheel velocities were observed during the dirt terrain, reflecting the UTV adaptation to the uneven surface. \Revised{Additionally, the performance of the proposed human leader pose recognition subsystem is evaluated in the final part of the experiment (from 65 to 80 seconds), where the UTV first stops when the leader assumes a crouched position and then moves backward, i.e. increased along-track error, when the leader raised left hand (from 65 to 70 seconds), and moves forward, i.e. decreased along-track error,  when the leader raised right hand (from 73 to 80 seconds).}
    
    %%%%%%%%%%%%%%%%%%%%%%%%%%%%%%%%%%%%%%
%%%%%%%%%%%%%%%%%%%%%%%%%%%%%%%%%%%%%%
%%%%%%%%%%%%%%%%%%%%%%%%%%%%%%%%%%%%%%

\section{Conclusions}
\label{sec:Section6}

%%%%%%%%%%%%%%%%%%%%%%%%%%%%%%%%%%%%%%
%%%%%%%%%%%%%%%%%%%%%%%%%%%%%%%%%%%%%%
%%%%%%%%%%%%%%%%%%%%%%%%%%%%%%%%%%%%%%

In this paper, a design of ADRC for the UTVs leader-follower control task was presented. The research aimed to address the challenges associated with the rapidly changing dynamics of the leader movement and potential measurement noise, thereby improving the tracking capabilities and adaptability of UTVs in real-world scenarios.
Through numerical simulations, the performance of the proposed ADRC approach was compared to that of conventional PI/PID control approaches. The results indicated that the ADRC controllers outperform the conventional controllers, particularly in scenarios involving slippage disturbances and measurement noise.
To experimentally validate the effectiveness of the proposed leader-follower control solution, a comprehensive experimental setup, which involved a laboratory UTV equipped with camera and laser sensors for track error estimation, as well as a subsystem for human leader pose recognition, was developed. The experimental results demonstrated that the UTV under ADRC control exhibited enhanced tracking capabilities, high robustness against slippage disturbances, and large adaptability to changing terrain conditions.

    %%%%%%%%%%%%%%%%%%%%%%%%%%%%%%%%%%%%%%
%%%%%%%%%%%%%%%%%%%%%%%%%%%%%%%%%%%%%%
%%%%%%%%%%%%%%%%%%%%%%%%%%%%%%%%%%%%%%

\section*{Acknowledgments}

%%%%%%%%%%%%%%%%%%%%%%%%%%%%%%%%%%%%%%
%%%%%%%%%%%%%%%%%%%%%%%%%%%%%%%%%%%%%%
%%%%%%%%%%%%%%%%%%%%%%%%%%%%%%%%%%%%%%

\begin{itemize}
    \item This research was supported by the University of Defence in Belgrade, Serbia, under grant no. VA-TT/1/21-23.
    \item The authors declare no competing interests.
\end{itemize}

%%%%%%%%%%%%%%%%%%%%%%%%%%%%%%%%%%%%%%
%%%%%%%%%%%%%%%%%%%%%%%%%%%%%%%%%%%%%%
%%%%%%%%%%%%%%%%%%%%%%%%%%%%%%%%%%%%%%

\appendix
%\section*{Appendix}

%%%%%%%%%%%%%%%%%%%%%%%%%%%%%%%%%%%%%%
%%%%%%%%%%%%%%%%%%%%%%%%%%%%%%%%%%%%%%
%%%%%%%%%%%%%%%%%%%%%%%%%%%%%%%%%%%%%%

%%%%%%%%%%%%%%%%%%%%%%%%%%%%%%%%%%%%%%
\section{Discrete-time transfer function of lateral ADRC controller}
%%%%%%%%%%%%%%%%%%%%%%%%%%%%%%%%%%%%%%

\begin{equation}
    G_\mathrm{FB_L}(z)=\frac{\beta_{10}+\beta_{11}z^{-1}+\beta_{12}z^{-2}}{1+\alpha_{11}z^{-1}+\alpha_{12}z^{-2}}\cdot\frac{1}{(1-z^{-1})},
    \label{Discrete-time-lateral}
\end{equation}
with:
%\shortstack{  $\beta_{10} =  \frac{1}{b_0T_s^2}\bigg[\frac{1}{4}\cdot(1+z_\mathrm{CL_l})^2\cdot(1+z_\mathrm{ESO_l})^3 -2\cdot(z_\mathrm{CL_l}^2\cdot z_\mathrm{ESO_l}^3$ \\
%$+2z_\mathrm{CL_l}+3z_\mathrm{ESO_l}-2)\bigg]$} 
\begin{align*}
    \alpha_{11} & = \frac{-1}{8}\cdot(1+z_\mathrm{CL_l})^2\cdot (1+z_\mathrm{ESO_1})^3+z_\mathrm{CL_l}^2\cdot z_\mathrm{ESO_l}^3+1,\\
    \alpha_{12} & =  z_\mathrm{CL_l}^2z_\mathrm{ESO_l}^3, \\
    \begin{split}
        \beta_{10} & = \frac{1}{b_0T_s^2}\bigg[\frac{1}{4}\cdot(1+z_\mathrm{CL_l})^2\cdot(1+z_\mathrm{ESO_l})^3 -2\cdot(z_\mathrm{CL_l}^2\cdot z_\mathrm{ESO_l}^3 \\
        & \quad+2z_\mathrm{CL_l}+3z_\mathrm{ESO_l}-2)\bigg],
    \end{split} 
    \\
    \begin{split}
        \beta_{11} & = \frac{1}{b_0T_s^2}\bigg[-(1+z_\mathrm{CL_l})^2\cdot (1+z_\mathrm{ESO_l})^3+2\cdot(1+z_\mathrm{CL_l})^2 \\
        & \qquad +6\cdot(z_\mathrm{CL_l}^2\cdot z_\mathrm{ESO_l}^3+2z_\mathrm{CL_l}\cdot z_\mathrm{ESO_l}+z_\mathrm{ESO_l}^2+z_\mathrm{ESO_l}-1)\bigg],
    \end{split} 
    \\
    \begin{split}
        \beta_{12} & = \frac{1}{b_0T_s^2}\bigg[-\frac{1}{4}\cdot(1+z_\mathrm{CL_l})^2\cdot (1+z_\mathrm{ESO_l})^3+2\cdot(-2z_\mathrm{CL_l}^2\cdot z_\mathrm{ESO_l}^3+\\
        & \qquad 3z_\mathrm{CL_l}^2\cdot z_\mathrm{ESO_l}^2 +2z_\mathrm{CL_l}\cdot z_\mathrm{ESO_l}^3+1)\bigg],
    \end{split}
\end{align*}
where $z_\mathrm{CL_1}=\exp{^{-\omega_\mathrm{CL_1}\cdot T_\mathrm{s}}}$ and $z_\mathrm{ESO_1}=\exp{^{-k_\mathrm{ESO_1}\cdot \omega_\mathrm{CL_1}\cdot T_\mathrm{s}}}$.

%%%%%%%%%%%%%%%%%%%%%%%%%%%%%%%%%%%%%%
\section{Discrete-time transfer functions of longitudinal ADRC controller}
%%%%%%%%%%%%%%%%%%%%%%%%%%%%%%%%%%%%%%

\begin{equation}
    G_\mathrm{FB_{v}}(z)=\frac{\beta_{20}+\beta_{21}z^{-1}}{1+\alpha_{21}z^{-1}}\cdot\frac{1}{(1-z^{-1})},
    \label{Discrete-time-longitudinal-FB}
\end{equation}
\begin{equation}
    G_\mathrm{PF_{v}}(z)=\frac{1}{\beta_{20}}\frac{\gamma_{20}+\gamma_{21}z^{-1}+\gamma_{22}z^{-2}}{1+\frac{1}{\beta_{20}}(\beta_{21}z^{-1})},
    \label{Discrete-time-longitudinal-PF}
\end{equation}
with:
\begin{align*}
    \alpha_{21} & = -2z_\mathrm{CL_v}\cdot z_\mathrm{ESO_v}^2, \\
    \beta_{20} & = \frac{1}{b_0T_s}[z_\mathrm{CL_v}\cdot z_\mathrm{ESO_v}^2-2z_\mathrm{ESO_v}-z_\mathrm{CL_v}+2], \\
    \beta_{21} & = \frac{1}{b_0T_s}[2z_\mathrm{CL_v}\cdot z_\mathrm{ESO_v}-2z_\mathrm{CL_v}\cdot z_\mathrm{ESO_v}^2+z_\mathrm{ESO_v}^2-1], \\
    \gamma_{20} & = \frac{1-z_\mathrm{CL_v}}{b_0T_s}, \\
    \gamma_{21} & = -2z_\mathrm{ESO_v}\cdot\frac{(1-z_\mathrm{CL_v})}{b_0T_s}, \\
    \gamma_{22} & = z_\mathrm{ESO_v}^2\cdot\frac{(1-z_\mathrm{CL_v})}{b_0T_s},
\end{align*}
where $z_\mathrm{CL_v}=\exp{^{-\omega_\mathrm{CL_v}\cdot T_\mathrm{s}}}$ and $z_\mathrm{ESO_v}=\exp{^{-k_\mathrm{ESO_v}\cdot \omega_\mathrm{CL_v}\cdot T_\mathrm{s}}}$.

 \bibliographystyle{elsarticle-num} 
 \bibliography{cas-refs}

\end{document}